\documentclass[twocolumn,aps,showpacs,prb]{revtex4-1}

\usepackage{graphicx}
\usepackage{epstopdf}
\usepackage{amsmath}
\usepackage{multirow}

\begin{document}

\title{First-principles study on the electron-phonon coupling and magnetoresistance of LaBi under pressure}

\author{Jian-Feng Zhang$^{1}$}
\author{Peng-Jie Guo$^{2}$}
\author{Miao Gao$^{3}$}
\author{Kai Liu$^{1}$}\email{kliu@ruc.edu.cn}
\author{Zhong-Yi Lu$^{1}$}\email{zlu@ruc.edu.cn}

\affiliation{$^{1}$Department of Physics and Beijing Key Laboratory of Opto-electronic Functional Materials $\&$ Micro-nano Devices, Renmin University of China, Beijing 100872, China}

\affiliation{$^{2}$Songshan Lake Materials Laboratory, Dongguan, Guangdong 523808, China}

\affiliation{$^{3}$Department of Microelectronics Science and Engineering, School of Physical Science and Technology, Ningbo University, Zhejiang 315211, China}

\date{\today}

\begin{abstract}

The extremely large magnetoresistance (XMR) material LaBi was reported to become superconducting under pressure accompanying with suppressed magnetoresistance. However, the underlying mechanism is unclear. By using first-principles electronic structure calculations in combination with a semiclassical model, we have studied the electron-phonon coupling and magnetoresistance of LaBi in the pressure range from 0 to 18 GPa. Our calculations show that LaBi undergoes a structural phase transition from a face-centered cubic lattice to a primitive tetragonal lattice at $\sim$7 GPa, verifying previous experimental results. Meanwhile, LaBi remains topologically nontrivial across the structural transition. Under all pressures that we have studied, the phonon-mediated mechanism based on the weak electron-phonon coupling cannot account for the observed superconductivity in LaBi, and the calculated magnetoresistance for LaBi does not show a suppression. The distinct difference between our calculations and experimental observations suggests either the existence of extra Bi impurities in the real LaBi compound or the possibility of other unknown mechanism.

\end{abstract}

\pacs{}

\maketitle

\section{INTRODUCTION}

Searching for new superconducting materials and exploring the related superconducting mechanism have long been the key issues in the study of superconductivity. While the conventional superconductivity can be well understood in the framework of electron-phonon coupling (EPC) according to Bardeen-Cooper-Schrieffer (BCS) theory \cite{bcs}, the unconventional superconductivity, which was found in cuprates \cite{ED1994,AD2003,PAL2006,RM1990}, iron-based superconductors \cite{LaOFeAsHH,LaOFeAsJMD,BaKFeAs,FeSe,Alireza08,Kimber09}, heavy-fermion compounds
\cite{GRS1984,VJE1994,NDM1998,MS1991} etc, is widely believed to correlate with spin fluctuations \cite{LaOFeAsJMD,GB1994}. Recently, a new type of materials, which show extremely large magnetoresistance (XMR) around $10^4\%$ to $10^7\%$ at ambient pressure \cite{Cava2014,Tafti2016,LeiHC2016}, can develop superconductivity under pressure \cite{Pan15,Lu2016,Tafti2017}. The emergence of superconductivity accompanying with suppressed magnetoresistance in these XMR materials is analogous to the one in unconventional superconductors, where the superconductivity is on the border of long-range magnetic orders. This novel phenomenon in the XMR materials kindles our interest to investigate the underlying superconducting mechanism, which may provide a reference for the unconventional superconductivity.

Among the XMR materials, lanthanum monopnictides LaSb and LaBi, which demonstrate similar XMR effect but distinct topological properties, have attracted intensive attention \cite{FuL2015,Tafti2016,GuoPJ2016,LeiHC2016,Claudia2016,DingH2016,AKaminski2016,DLFeng2016,Tafti2017,GuoPJ2017,Nayak2017,Lou2017,Kumar2017,Singha2017,Oinuma2017,Tafti2016pnas}, being model materials. 
LaBi has a face-centered cubic (fcc) lattice and its XMR can reach $10^4\%$ at ambient pressure \cite{LeiHC2016}. Experimentally, under 3.5 GPa, the XMR of LaBi is suppressed and meanwhile the superconductivity emerges with a transition temperature $T_c$ of $\sim$4 K \cite{Tafti2017}. With further increasing pressure, $T_c$ first rises to $\sim$6.5 K around 7 GPa, then decreases gradually to $\sim$5.5 K until a structural phase transition to a primitive tetragonal (pt) lattice at 11 GPa, followed by a jump of $T_c$ to $\sim$8 K. Beyond 11 GPa, $T_c$ keeps decreasing with pressure \cite{Tafti2017}. Previous studies on a two-dimensional XMR material WTe$_2$ \cite{Cava2014} suggested that the electron-phonon coupling is responsible for the observed superconductivity in pressed WTe$_2$ \cite{Kang15,Pan15,Lu2016}. In contrast, the origin of superconductivity in the three-dimensional XMR material LaBi under pressure is unresolved \cite{Tafti2017}.

In this work, we have studied the evolutions of crystal structure, electronic structure, phonon spectrum, electron-phonon coupling, and magnetoresistance of LaBi with pressure by using first-principles calculations. We find that no matter whether LaBi is in the fcc structure at low pressure or in the pt structure at high pressure, the calculated $T_c$s from the EPC all approach to 0 K, indicating that the EPC mechanism cannot account for the observed superconductivity. Moreover, in comparison with the suppressed magnetoresistance in previous transport measurement, our calculated carrier densities and mobilities suggest stable magnetoresistance ($\sim$10$^4\%$) under pressure. We then discuss the possible reasons for the discrepancy between our calculations and previous experimental observations on LaBi.

\section{COMPUTATIONAL DETAILS}

We investigated the electronic structures and phonon spectra of LaBi under three representative pressures of 0, 6, and 15 GPa based on the density functional theory (DFT) \cite{dft1, dft2} and density functional perturbation theory (DFPT) \cite{dfpt,dfptreview} calculations as implemented in the Quantum ESPRESSO (QE) package \cite{pwscf}. The interactions between electrons and nuclei were described by the norm-conserving pseudopotentials \cite{ncpp}. The valence electron configurations are $5s^2 5p^6 5d^1 6s^2$ for La and $5d^{10} 6s^2 6p^3$ for Bi. For the exchange-correlation functional, the generalized gradient approximation (GGA) of Perdew-Burke-Ernzerhof (PBE) \cite{PBE} type was adopted. The kinetic energy cutoff of plane-wave basis was set to be 80 Ry. The Gaussian smearing method with a width of 0.004 Ry was employed for the Fermi surface broadening. The spin-orbit coupling (SOC) effect was included as La and Bi are heavy elements. In structural optimization, both lattice constants and internal atomic positions were fully relaxed until the forces on all atoms were smaller than 0.0002 Ry/Bohr.

The superconducting $T_c$ was studied based on the EPC theory as implemented in the EPW package \cite{epw}, which uses the maximally localized Wannier functions \cite{mlwf} and interfaces with the QE \cite{pwscf}. We took the 8$\times$8$\times$8 {\bf k}-mesh and 4$\times$4$\times$4 {\bf q}-mesh as coarse grids and then interpolated to the 48$\times$48$\times$48 {\bf k}-mesh and 16$\times$16$\times$16 {\bf q}-mesh as dense grids respectively. The EPC constant $\lambda$ can be calculated either by the summation of EPC constant $\lambda_{{\bf q}\nu}$ in the whole Brillouin zone (BZ) for all phonon modes or by the integral of Eliashberg spectral function $\alpha^2F(\omega)$ as below \cite{Eliashberg},

\begin{equation}
\lambda=\sum_{{\bf q}\nu}\lambda_{{\bf q}\nu}=2\int{\frac{\alpha^2F(\omega)}{\omega}d\omega}.
\end{equation}
The Eliashberg spectral function is defined as\cite{Eliashberg},
\begin{equation}
\alpha^2F(\omega)=\frac{1}{2{\pi}N(\varepsilon_F)}\sum_{{\bf q}\nu}\delta(\omega-\omega_{{\bf q}\nu})\frac{\gamma_{{\bf q}\nu}}{\hbar\omega_{{\bf q}\nu}},
\end{equation}
where $N(\varepsilon_F)$ is the density of states (DOS) at Fermi level $\varepsilon_F$, $\omega_{{\bf q}\nu}$ is the frequency of the $\nu$-th phonon mode at wave vector {\bf q}, and $\gamma_{{\bf q}\nu}$ is the phonon linewidth\cite{Eliashberg},
\begin{equation}
\gamma_{{\bf q}\nu}=2\pi\omega_{{\bf q}\nu}\sum_{{\bf k}nn'}|g_{{\bf k+q}n',{\bf k}n}^{{\bf q}\nu}|^2\delta(\varepsilon_{{\bf k}n}-\varepsilon_F)\delta(\varepsilon_{{\bf k+q}n'}-\varepsilon_F),
\end{equation}
in which $g_{{\bf k+q}n',{\bf k}n}^{{\bf q}\nu}$ is the electron-phonon coupling matrix element.
The superconducting transition temperature $T_c$ can be predicted by substituting the EPC constant $\lambda$ into the McMillan-Allen-Dynes formula \cite{McMillan1, McMillan2},
\begin{equation}
T_c=\frac{\omega_{log}}{1.2}exp[\frac{-1.04(1+\lambda)}{\lambda(1-0.62\mu^*)-\mu^*}],
\end{equation}
where $\mu^*$ is the effective screened Coulomb repulsion constant. In our calculation, $\mu^*$ was set to 0.1, between the widely-used empirical values of 0.08 and 0.15\cite{mustar1,mustar2}. The logarithmic average of the Eliashberg spectral function $\omega_{log}$ is defined as\cite{McMillan1, McMillan2},
\begin{equation}
\omega_{log}=exp[\frac{2}{\lambda}\int{\frac{d\omega}{\omega}\alpha^2F(\omega){ln}(\omega)}].
\end{equation}

\begin{figure}[!b]
\includegraphics[angle=0,scale=0.29]{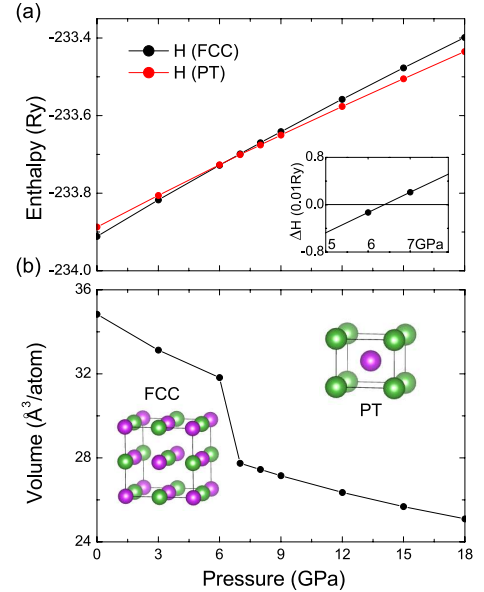}
\caption{(Color online) (a) The pressure-dependent enthalpy of LaBi in the face-centered cubic (fcc) and the primitive tetragonal (pt) phases. The inset shows the enthalpy difference between these two phases around the critical transition point: $\Delta$H = H(fcc) - H(pt). (b) The evolution of cell volumes (in unit of {\AA}$^3$/atom) of LaBi in the low-enthalpy lattice structure with increasing pressure. The inset shows the crystal structures of fcc and pt lattices, where the green and purple balls represent La and Bi atoms, respectively.}
\label{fig1}
\end{figure}

The magnetoresistance (MR) of LaBi was studied based on the semiclassical two-band model \cite{twob1,twob2}. In the condition of perfect charge compensation, the formula of magnetoresistance can be reduced as,
\begin{equation}
\rm MR = \mu_e \mu_h B^2,
\end{equation}
where $\mu_e$ and $\mu_h$ are respectively the electron-type and hole-type carrier mobilities, and B is the magnetic field. The carrier mobilities were studied based on Boltzmann transport equation (BTE) with self-energy relaxation time approximation (SERTA), which was implemented in the EPW package\cite{epwmob}. As LaBi owns intrinsic carriers, we only considered the electron-phonon scattering and ignored the impurity scattering. With SERTA, the mobility takes the following simple form\cite{epwmob},
\begin{equation}
\mu_{e,\alpha\beta}=\frac{-e}{n_e\Omega}\sum_{n\in \rm CB}{\int{\frac{d\bf k}{\Omega_{\rm BZ}}\frac{\partial{f^{0}_{n\bf k}}}{\partial{\varepsilon_{n\bf k}}}v_{n\bf k,\alpha}v_{n\bf k,\beta}\tau^{0}_{n\bf k}}},
\end{equation}
where $n_e$ is the electron-type carrier density, $\Omega$ is the cell volume, $\Omega_{\rm BZ}$ is the BZ volume, $f^0_{n\bf k}$ is the Fermi-Dirac distribution, $v_{{n\bf k},\alpha}=\hbar^{-1}\partial\varepsilon_{n\bf k}/\partial{k_\alpha}$ is the band velocity, and $\tau^0$ is the relaxation time defined as\cite{epwmob},
\begin{equation}
\begin{split}
\frac{1}{\tau^{0}_{n\bf k}}=\frac{2\pi}{\hbar}\sum_{n'\nu}\int{\frac{d\bf q}{\Omega_{\rm BZ}}}|g_{{\bf k+q}n',{\bf k}n}^{{\bf q}\nu}|^2{\ \ \ \ \ \ \ \ \ \ \ \ \ \ \ \ \ \ \ \ \ }\\
[(1-f^0_{n'\bf k+\bf q}+n_{{\bf q}\nu})\delta(\varepsilon_{n\bf k}-\varepsilon_{n'\bf k+\bf q}-\hbar\omega_{\bf q\nu})\\
+(f^0_{n'\bf k+\bf q}+n_{\bf q\nu})\delta(\varepsilon_{n\bf k}-\varepsilon_{n'\bf k+\bf q}+\hbar\omega_{\bf q\nu})]{\ \ \ }
\end{split},
\end{equation}
where $n_{\bf q\nu}$ is the Bose-Einstein distribution.

\begin{figure*}[tb]
\includegraphics[angle=0,scale=0.65]{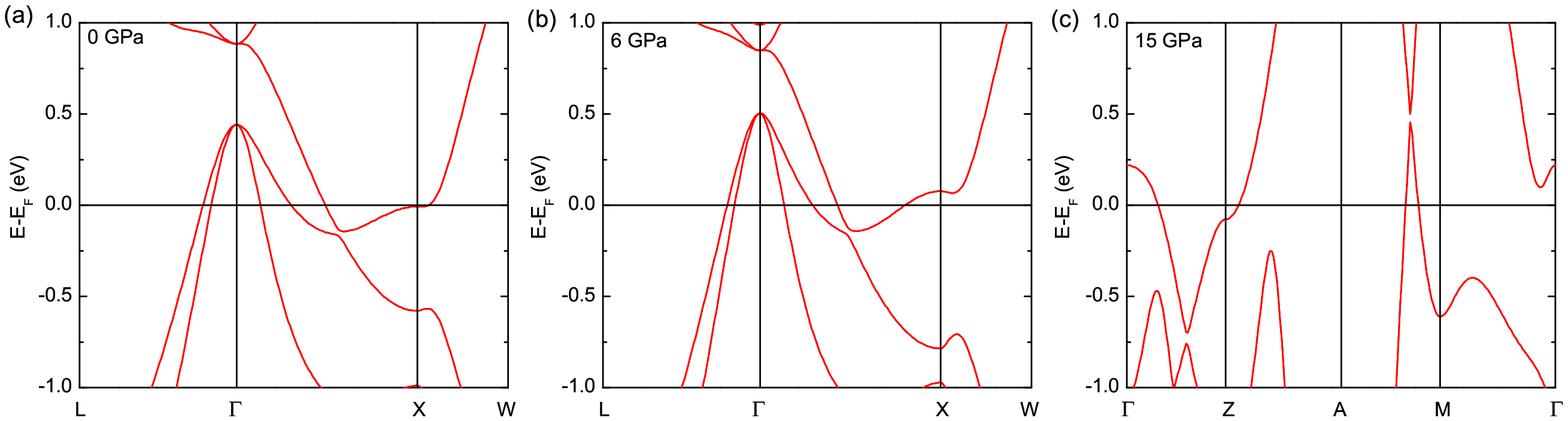}
\caption{(Color online) Electronic band structures of LaBi at (a) 0 GPa in fcc lattice, (b) 6 GPa in fcc lattice, and (c) 15 GPa in pt lattice.}
\label{fig2}
\end{figure*}

\begin{figure*}[tb]
\includegraphics[angle=0,scale=0.65]{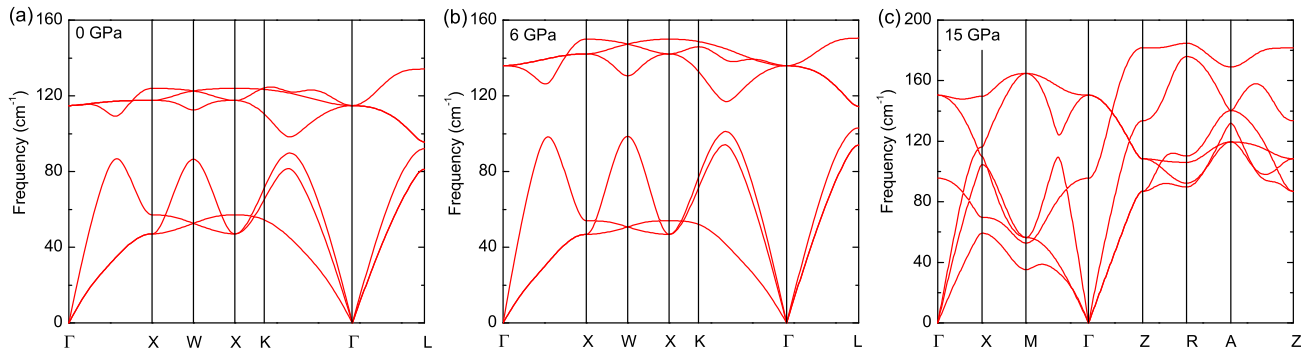}
\caption{(Color online) Phonon dispersions of LaBi at (a) 0 GPa in fcc lattice, (b) 6 GPa in fcc lattice, and (c) 15 GPa in pt lattice.}
\label{fig3}
\end{figure*}

\section{Results and Analysis}

First of all, we have investigated the crystal structures of LaBi under different pressures. In experiment \cite{Tafti2017}, a structural phase transition from the fcc lattice to the pt lattice was observed around 11 GPa. We have thus studied the enthalpies of LaBi in the fcc and pt structures within a pressure range (0-18 GPa) covering the above pressure [Fig. 1(a)]. The calculated enthalpy of the fcc lattice is lower than that of the pt lattice at low pressure until a reversion takes place around 7 GPa. The cell volumes of LaBi in the low-enthalpy crystal structures under corresponding pressures are shown in Fig. 1(b). The calculated cell volumes (lattice constants) are all smaller than the experimental values by about 1.8\% (0.6\%) in the whole pressure range except those around the structural transition point, indicating the good agreement between our calculations and previous measurements\cite{Tafti2017}. Around the structural phase transition, there is a sudden reduction of cell volume. The calculated transition pressure is about 7 GPa, which is lower than the experimental value (11 GPa)\cite{Tafti2017}. This difference may be attributed to the temperature effect, the anharmonic effect, and/or the impurities in real synthesized compound (we will discuss this point later). Overall, these structural features of LaBi from our calculations verify those observed in previous experiments\cite{Tafti2017}.

\begin{figure*}[tb]
\includegraphics[angle=0,scale=0.62]{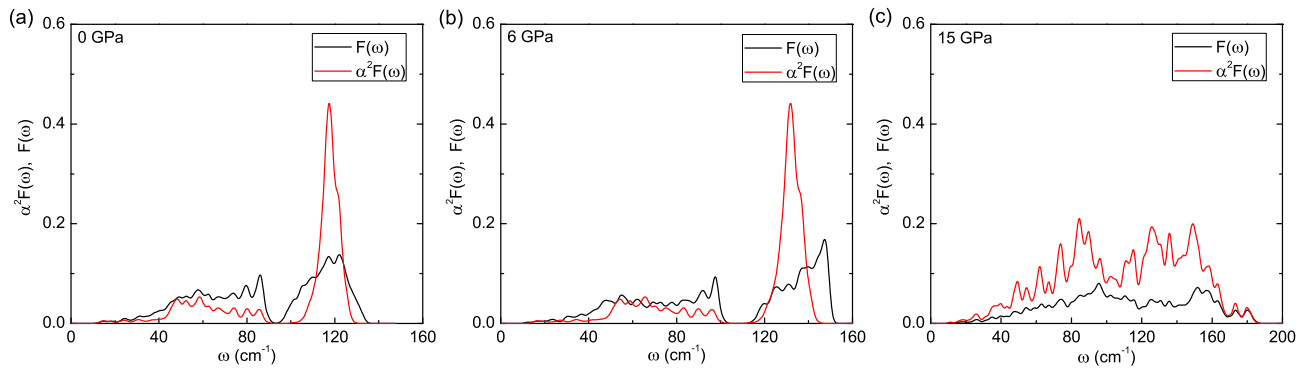}
\caption{(Color online) Phonon density of states $F(\omega)$ (black line) and Eliashberg spectral function $\alpha^2F(\omega)$ (red line) at (a) 0, (b) 6, and (c) 15 GPa.}
\label{fig4}
\end{figure*}

Based on the equilibrium lattices under various pressures, the electronic band structures of LaBi (Fig. 2) were further studied. We mainly focused on three pressures\cite{Tafti2017}: (1) the ambient pressure (0 GPa) at which no superconductivity is found in experiment; (2) the pressure with optimal $T_c$ experimentally for LaBi in fcc lattice (6 GPa); and (3) the pressure for LaBi in pt lattice (15 GPa). A comparison between the band structures under 0 and 6 GPa [Figs. 2(a) and 2(b)] shows that the hole-like pockets around the $\Gamma$ point vary slowly with the increasing pressure, while the electron-like pocket around the $X$ point exhibits dramatic changes. To be specific, the band at the $X$ point shows a plateau below the Fermi level at 0 GPa, but it shifts above the Fermi level at 6 GPa [Fig. 2(b)], indicating a Lifshitz transition. Nevertheless, the reservation of band inversion around the $X$ point at 6 GPa makes LaBi maintain its nontrivial topological property, as verified by the calculated wavefunction parity product at the time-reversal invariant points. For the pt structure at 15 GPa, LaBi also holds the nontrivial property, yet its density of states at the Fermi level increases by a factor of two (Table I).

The calculated phonon spectra of LaBi under these pressures are shown in Fig. 3. As the pressure increases, most phonon frequencies rise up due to the strengthened atomic bondings. There is a gap between the acoustic and optical branches at low pressures [Figs. 3(a) and 3(b)], which diminishes after the structural phase transition [Fig. 3(c)]. The phonon density of states $F(\omega)$ and the corresponding Eliashberg spectral function $\alpha^2F(\omega)$ are plotted in Fig. 4. In the whole frequency range, the intensities of the Eliashberg spectral function rarely exceed 0.4, mostly below 0.2. This indicates that the electron-phonon coupling is very weak in LaBi.
With the knowledge of the Eliashberg spectral function, the total EPC constant $\lambda$ can be calculated and then the superconducting transition temperature $T_c$ can be obtained based on the McMillan-Allen-Dynes formula. As listed in Table I, the calculated $T_c$s of LaBi in the fcc lattice under 0 and 6 GPa all approach to 0 K, revealing the absence of EPC-derived superconductivity and the weak influence of the pressure. After the structural phase transition, the superconducting $T_c$ increases slightly (0.12 K). However, it is still much lower than the experimental value ($\sim$8 K) \cite{Tafti2017}. Even an adjustment of $\mu^*$ to 0.0, which yields the highest $T_c$ (about 1.4 K) according to Eq. (4), does not change our conclusion.
The above results suggest that the experimentally observed superconductivity in pressed LaBi, both in fcc and pt lattices, does not originate from the conventional EPC mechanism.

\begin{figure*}[tb]
\includegraphics[angle=0,scale=0.65]{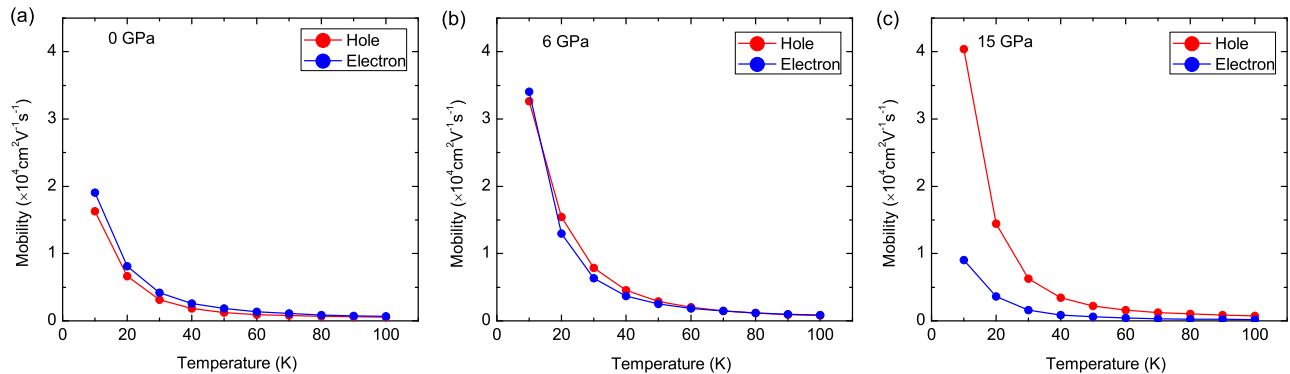}
\caption{(Color online) The temperature-dependent hole-type and electron-type carrier mobilities at (a) 0, (b) 6, and (c) 15 GPa.}
\label{fig5}
\end{figure*}

\begin{table}[!b]
\caption{The calculated electronic density of states (DOS) at the Fermi level $N(E_f)$ (in unit of states/eV), the logarithmic average of Eliashberg spectral function $\omega_{log}$ (in unit of cm$^{-1}$), the electron-phonon coupling $\lambda$, the superconducting $T_c$ (in unit of K), the carrier concentrations $n_{e,h}$ (in unit of 10$^{20}$cm$^{-3}$), the carrier mobilities $\mu_{e,h}$ at 10 K (in unit of 10$^{4}$cm$^{2}$V$^{-1}$s$^{-1}$), and the magnetoresistance (MR) (in unit of 10$^4\%$) of LaBi at 9 T and 10 K under 0, 6, and 15 GPa, respectively.}
\begin{center}
\begin{tabular*}{8cm}{@{\extracolsep{\fill}} cccc}
\hline \hline
Pressure & 0 & 6 & 15 \\
\hline
$N(E_f)$ & 0.48 & 0.38 & 0.99 \\
$\omega_{log}$ & 78.0 & 80.3 & 83.1 \\
$\lambda$ & 0.124 & 0.132 & 0.325 \\
$T_c$ & 0.00 & 0.00 & 0.12 \\
$n_{e,h}$ & 4.0 & 3.9 & 19.8 \\
$\mu_e$ & 1.9 & 3.4 & 0.9 \\
$\mu_h$ & 1.6 & 3.2 & 4.0 \\
MR & 2.5 & 9.0 & 3.0 \\
\hline
\hline
\end{tabular*}
\end{center}
\end{table}

In previous experiment \cite{Tafti2017}, the pressure not only induces superconductivity in LaBi but also completely suppresses its magnetoresistance. The calculated carrier densities in Table I indicate that  LaBi is in good electron-hole compensation. According to the semiclassical two-band model, the magnetoresistance of a charge-compensated semimetal is reduced to $\rm MR = \mu_e\mu_hB^2$, which is merely determined by the product of carrier mobilities and the square of magnetic field. The calculated carrier mobilities ($\mu_{e,h}$) of LaBi are shown in Fig. 5, which decrease quickly with increasing temperature. In particular, the calculated carrier mobilities of LaBi in fcc lattice at 0 GPa [Fig. 5(a)] agree quite well with the previous measured values\cite{LeiHC2016}. For the same fcc lattice of LaBi at 6 GPa, both $\mu_{h}$ and $\mu_{e}$ take obvious increments in comparison with those at 0 GPa, resulting in a large MR. This can be understood from the fact that the pressure can broaden the bands and increase the Fermi velocity $v_{n\bf k}(\varepsilon_F)$, so as to enhance the carrier mobilities $\mu_{e,h}$ [Eq. (7)]. On the other hand, for the pt structure of LaBi under 15 GPa, the product of $\mu_{e}$ and $\mu_{h}$ reduces due to the dramatic increase of carrier concentrations [Table I and Eq. (7)]. As a result, the MR decreases at 15 GPa. Although the calculated MR first increases with pressure in fcc lattice and then decreases after the structural phase transition to pt lattice, it still maintains the order of $10^4\%$ (Table I), which disagrees with the rapid suppression in experimental observation\cite{Tafti2017}.

\section{Discussion and Summary}

The pressure is a clean approach for modulating the material properties.
Previously, the pressure-induced superconductivities have been observed in a series of  materials, such as sulfur hydride \cite{Drozdov15}, lanthanum superhydride \cite{lah1,lah2}, BaFe$_2$As$_2$ \cite{Alireza08,Kimber09}, etc. For sulfur hydride and lanthanum superhydride, whose respective $T_c$ can reach 203 and 260 K under extremely high pressures, a prominent isotope shift of $T_c$ indicates the EPC mechanism namely the conventional superconductivity \cite{Drozdov15, maprl}. As to the undoped BaFe$_2$As$_2$, the pressure-destabilized antiferromagnetic order in ground state may lead to the emergence of the unconventional superconductivity \cite{LaOFeAsJMD,Kimber09}. Here, for the pressed LaBi, our calculations demonstrate that the EPC alone cannot account for its superconductivity found in experiment \cite{Tafti2017}. Moreover, since no magnetism has been observed in LaBi, the magnetic (spin) fluctuations are impossible to take part. As a result, there may be other novel mechanism involved in the superconductivity of LaBi under pressure.

Beyond the opinion of inherent superconductivity in LaBi under pressure, there is also one possibility that the impurity in LaBi may play a role \cite{Tafti2017}. For example, the elemental Bi crystal is not superconducting at ambient pressure, but it can transform into Bi-III phase at 2.7 GPa with a $T_c$ about 8 K and then enter another phase at $\sim$8 GPa with a jump of $T_c$ \cite{LiYF2017}. In view of the phase diagrams\cite{Tafti2017}, the similarities in the $T_c$ values and the critical pressures between elemental Bi and LaBi suggest the possibility of extra Bi impurity in LaBi. In addition to the Bi impurity, an intermetallic compound LaBi$_3$, recently synthesized from Bi and LaBi under pressure, also shows a comparable $T_c$ of 7.3 K \cite{Lyo2016}. This provides one more possibility to observe superconductivity in realistic LaBi compound. In fact, the existence of impurities may also bring about the aforementioned difference in structural transition pressure between our calculations and previous measurement\cite{Tafti2017}.

On the other hand, for the suppression of MR under pressure, which was observed in experiment\cite{Tafti2017} but was not reproduced by our calculations for pure LaBi, it may be understood by a derivation from the emergent superconductivity in realistic LaBi compound, i.e., the enhanced electron-phonon coupling under pressure that induces superconductivity via Bi impurities or intermetallic compound LaBi$_3$ will also bring strong scattering of transport carriers. This will dramatically influence carrier mobilities and then suppress MR, as deduced from Eqs. (1)-(8), where $\lambda$ ($\mu_{e,h}$) have positive (negative) relations with $\gamma_{{\bf q}\nu}$. So the key difference between our calculations and previous experiment is likely related to the different EPC strengths we obtained for pure LaBi in calculations and those in real synthesized compound containing impurities\cite{Tafti2017}.

In summary, we have studied the evolution of crystal structure, electronic/phonon band structure, superconducting property, and magnetoresistance of LaBi with pressure by using first-principles calculations. Our calculations verify a pressure-induced structural phase transition from fcc lattice to pt lattice in previous experiment. Nevertheless, in both lattice structures of LaBi, the calculated superconducting transition temperatures resulting from the EPC are far below the measured values, which means that the conventional EPC mechanism cannot explain the observed superconductivity in pressed LaBi. With the compensated carrier densities and the high carrier mobilities, our calculated magnetoresistance of LaBi does not show obvious suppression under pressure, which disagrees with the experimental observation either. Considering these substantial differences, we suggest the possibility that either Bi impurity (or intermetallic compound) or other novel mechanism may be responsible for the emergent superconductivity and the suppressed MR of LaBi under pressure, which waits for further experimental examination.

\begin{acknowledgments}

We thank H.-C. Lei and X.-Q. Lu for helpful discussions. This work was supported by the National Natural Science Foundation of China (Grants No. 11774424, No. 11774422, and No. 11974194), the National Key R\&D Program of China (Grants No. 2017YFA0302903 and No. 2019YFA0308603), the CAS Interdisciplinary Innovation Team, the Fundamental Research Funds for the Central Universities, and the Research Funds of Renmin University of China (Grant No. 19XNLG13). Computational resources were provided by the Physical Laboratory of High Performance Computing at Renmin University of China.

\end{acknowledgments}

\end{document}